\documentclass{article}
\usepackage{amssymb}
\usepackage{amsmath}
\usepackage{xypic}
\usepackage{graphicx}  
\usepackage{epstopdf}
\usepackage[colorlinks=true]{hyperref}

\newtheorem{theorem}{Theorem}[section]
\newtheorem{lemma}[theorem]{Lemma}
\newtheorem{remark}{\it Remark\/}
\newtheorem{proposition}[theorem]{Proposition}

\title{On the Poincar\'e's generating function and the symplectic mid-point rule}

\author{Hugo Jim\'enez-P\'erez, Jean-Pierre Vilotte and Barbara Romanowicz}

\newcommand{\II}{{\mathcal I}}
\newcommand{\JJ}{{\mathcal J}}
\newcommand{\KK}{{\mathcal K}}
\newcommand{\bb}{{\bf b}}
\newcommand{\bS}{{\bf S}}
\newcommand{\bR}{{\bf R}}
\newcommand{\bG}{{\bf G}}
\newcommand{\bA}{{\bf A}}
\newcommand{\bq}{{\bf q}}
\newcommand{\bp}{{\bf p}}
\newcommand{\bQ}{{\bf Q}}
\newcommand{\bP}{{\bf P}}

\newcommand{\bx}{{\bf x}}

\newcommand{\bz}{{\bf z}}

\newcommand{\mP}{{\mathcal P}}

\newcommand{\fiz}{ { \textstyle{ \frac{1}{2} } } }

\begin{document}
\maketitle

\begin{abstract}

The use of Liouvillian forms to obtain symplectic maps for constructing
numerical integrators is a natural alternative to 
the method of generating functions, and provides a 
deeper understanding of the geometry of this procedure. 
Using Liouvillian forms we study the generating function introduced by
Poincar\'e (1899) and its associated symplectic map. 
We show that in this framework, 
Poincar\'e's generating function does not correspond to the  
symplectic mid-point rule, but to the identity map.
We give an interpretation of this result based on the original 
framework constructed by Poincar\'e.

\end{abstract}

%

\section{Introduction}

From the available tools for constructing symplectic maps,
the method of generating functions has been a cornerstone to understand the 
links between the geometry and topology of the phase space of Hamiltonian 
mechanical systems. 
In his famous 
\emph{Les m\'ethodes nouvelles de la m\'ecanique c\'eleste}  \cite{Poi99}, Poincar\'e 
develops the theory of integral invariants with applications to the study 
of periodic orbits in celestial mechanics. 
Poincar\'e constructed a locally exact differential 1-form defined on closed orbits 
with  prescribed fixed period $T>0$, such that its exterior differential
gives the canonical symplectic form on the phase space. This 1-form
is the differential of a function known as the Poincar\'e's generating function \cite{Wei72,KQ10}
which is defined locally on a Lagrangian submanifold where it concides with a 
Liouvillian form. Since the orbit was periodic
he considered a  section\footnote{The term \emph{section} has a different 
meanning here than in holomogical algebra or vector bundles.}
(the Poincar\'e's section) and a non-trivial map defined 
on  such a  section (the Poincar\'e's map) such that the periodic orbit 
corresponds to a fixed point of the map. The imposed condition for the fixed point
was that the first-return map must be \emph{non-reversing} \cite{Poi99,Wei72}. 

From the numerical point of view, generating functions are used to construct 
numerical algorithms preserving the main geometrical properties of 
the phase space, in particular 
the symplectic form, naming the numerical algorithms as 
\emph{symplectic integrators}. 
Symplectic integrators 
are just the algorithmic numerical realization 
of symplectic maps close to the identity map, under some particular constraints. 
In the second half of the 80's, the construction of symplectic integrators using generating functions was 
systematically studied by Feng Kang and co-workers \cite{Kang85a, Kang85b, KG88, KHMD89}.  
In those papers, the Poincar\'e's generating function was
associated to the symplectic mid-point integrator. However,
there is no formal proof or construction of this correspondence.
Recently, an alternative method for constructing symplectic maps has been developed using 
Liouvillian forms \cite{Jim15a}.
This method 
was developped for working with
exterior differential forms in order to handle covariant objects. It becomes a natural 
alternative to the method of generating functions in the following way: 
A generating function $S$ defines a Lagrangian submanifold $\Lambda\subset M$, meanwhile a Liouvillian form $\theta$ defines a
codimension 1 coisotropic submanifold $C\subset M$, such that if $dS|_\Lambda=\theta|_\Lambda$
then $\Lambda\subset C$ and  $T\Lambda \subset TC\subset TM$ are vector sub-blundles. In this case we have 
$\ker\theta \subset \ker \pi^*dS$, for 
a suitable projection $\pi: M\to \Lambda$. Instead of solving the Hamilton-Jacobi equation 
for $\fiz \dim(M)$ local constants, we use the 1 dimensional kernel of $\theta$
for constructing a symplectic map, in fact we use the local Liouville vector field $Z$ (dual to 
Liouvillian form $\theta= i_Z\omega$) over the Hamiltonian flow for approximating the 
deviation of the numerical solution from the exact solution. 

In contrast with other methods for constructing symplectic integrators, 
in the method of Liouvillian forms we can approach the continuous flow of a generic 
autonomous Hamiltonian system by a classical result relating Hamiltonian and Liouville 
vector fields \cite{HZ12,MS17}. The application for sympelctic integrators is given in  \cite{Jim15a}. 

We can approach the method of Liouvillian forms using special symplectic manifolds \cite{Tul76,Tul77}. 
Alternatively, we can define a quaternionic or hypercomplex structure \cite{Boy88} on the product manifold, which produces transparent definitions, giving more 
information about the symplectic map and a better interpretation of its geometry.
The quaternionic structure induces three different symplectic forms, and consequently, three different
families of Liouvillian forms, one for each symplectic form (see Sec \ref{sec:sym}). 
Symplectic maps obtained in the linear approximation coincide with those found by Feng Kang and 
collaborators using generating functions and matrix algebra \cite{KQ10}. These maps were 
used by Kang as the input for the Hamilton-Jacobi equation, and they depend on a Hamiltonian 
matrix $b\in\mathbb M_{2n\times 2n}(\mathbb R)$ without any particular interpretation. 
In our case, we have a $(1,1)$-tensor, denoted by $\bb$, which is related with the way the solution curves 
in the phase space. This tensor comes from the symmetric part of the Liouvillian form.
When $\bb$ is a constant tensor, it becomes the matrix $b$ studied by Kang and collaborators. Moreover,
in this case we have solutions with constant curvature corresponding to the flow of 
quadratic Hamiltonian functions. 

A remarkable result is that we can construct symplectic integrators adapted to any well posed Hamiltonian problem,
since we can associate a Liouville vector field to any regular level hypersurface of a Hamiltonian 
function. This is a classical problem in the interface of contact and symplectic geometry \cite{HZ12}. 
By symplectic duality, we can associate a Liouvillian form to the system, and consequently a 
symplectic integrator. 
This association is locally defined on a prescribed hypersurface level and it defines a linear bundle 
whose dimension depends on the symmetries of the Hamiltonian.
If the Hamiltonian system have not other first integral than the Hamiltonian function,
then it is a real line bundle (dimension 1). In other case, we have a group of symmetries 
$G$ acting in a Hamiltonian way on the level surfaces and the bundle is related with the corresponding momentum map. 
This point of view has not been developped for the moment.

Liouvillian forms let us associate: 
1) Hamiltonian systems, 2) Liouville vector fields,
 and 3) symplectic maps. In particular, 
 we are interested in the relation between Liouvillian forms and symplectic maps.
In contrast to the claim found in the 
papers of Kang and his collaborators, our 
results associate the differential of the Poincar\'e's generating function to the identity map.
This is not a surprise since this is the original goal of Poincar\'e. Moreover, the
mid-point rule and the Poincar\'e's differential 
form belong to different families of minimizers of the action integral.
Indeed, the mid-point rule minimizes the action along a path with different 
fixed boundary points. Meanwhile Poincar\'e's differential form minimizes the 
action integral along a periodic closed path with prescribed fixed period
$T>0$, characterized on a Poincar\'e's section
by a fixed point \cite{Poi99}. 

The goal of this communication is to understand the structure of Poincar\'e's 
generating function in the canonical coordinates of the product manifold showing that 
its structure is completely different to the Liouvillian form producing the mid point rule.

\section{Symplectic integrators from Liouvillian forms}
\label{sec:sym} 
For more detail concerning this section, we refer the reader to \cite{Jim15a,Jim19a}.
Let $(M,\omega)$ be a $2n$-dimensional exact symplectic manifold
with local coordinates $\{q_i,p_i\}_{i=1}^n\in M$,
and $J_{2n}$ the canonical 
complex structure or canonical symplectic matrix given by 
\begin{eqnarray}
    J_{2n} = \left( 
    \begin{array}{c c }
        0_{n} & I_n\\
        -I_n & 0_n
    \end{array}
    \right), \qquad I_n,0_n\in \mathbb M_{n\times n}(\mathbb R).
    \label{eqn:J0}
\end{eqnarray}

A \emph{symplectomorphism} $\phi:(M_1,\omega_1)\to (M_2,\omega_2)$ is defined as
a diffeomorphism satisfying 
$\phi^*\omega_2=\omega_1$. Consider the product manifold $\mP =M_1\times M_2$
and two differential forms induced by the 
canonical projections $\pi_i:\mP\to M_i$, $i=1,2$ given by 
$\omega_{\ominus} = \pi_1^*\omega_1 - \pi_2^*\omega_2$ and $\theta_{\ominus} = \pi_1^*\theta_1 - \pi_2^*\theta_2$  .
The couple $(\mP,\omega_{\ominus}=d\theta_\ominus)$ is
an exact symplectic manifold of dimension $4n$ 
\cite{LM87}, where the graph of $\phi$ 
defined by the set 
$$\Gamma= \{\left(\bz,\phi(\bz)\right)|\bz\in M_1,\phi(\bz)\in M_2 \},$$ 
represents a Lagrangian submanifold $\Gamma\subset(\mP,\omega_\ominus)$. 
Given canonical coordinates on the factor manifolds $(M_1,\omega_1)$, $(M_2,\omega_2)$, and an Euclidean structure
$\langle\cdot,\cdot\rangle_{4n}$ on $T_m\mP$, $m\in\mP$, 
we define three complex structures 
\begin{eqnarray*}
     \II = \left( 
    \begin{array}{c c }
        0_{2n} & -I_{2n}\\
        I_{2n} & 0_{2n}
    \end{array}
    \right), \quad     
     J_\ominus = \left( 
    \begin{array}{c c }
        J_{2n} & 0_{2n}\\
        0_{2n} & J^T_{2n}
    \end{array}
    \right) \quad {\rm and}\quad
    \KK = \left( 
    \begin{array}{c c }
        0_{2n} & J_{2n} \\
        J_{2n} & 0_{2n}
    \end{array}
    \right),
\end{eqnarray*}
where $0_{2n},I_{2n},J_{2n}\in \mathbb M_{2n\times 2n}(\mathbb R)$.
The set $\{I_{4n},\II,\JJ,\KK\}\in End(T\mP)$  
induces an \emph{almost quaternionic} or \emph{almost hypercomplex structure}\footnote{We use $\II = -J_{4n}=J_{4n}^T$ for avoiding a 
negative sign in the quaternionic structure. The negative sign has been the source of different sign conventions 
between symplectic and complex geometries (see Remark 3.1.6 in \cite{MS17}).}, 
and three different symplectic structures on $\mP$ given by
$\omega_\II(\cdot,\cdot) = \langle\cdot, \II\cdot\rangle$,  
$\omega_\JJ(\cdot,\cdot) = \langle\cdot, \JJ\cdot\rangle$ and 
$\omega_\KK(\cdot,\cdot) = \langle\cdot, \KK\cdot\rangle$, where $\omega_\JJ\equiv\omega_\ominus$.

In this framework, the $2n$-dimensional submanifolds $\Lambda\stackrel{\jmath}{\hookrightarrow} \mP$ adapted for 
constructing symplectic maps are those which are Lagrangian with respect to 
$\omega_\II$ and $\omega_\JJ$ and symplectic with respect to $\omega_\KK$ \cite{Jim19a}.
The simplest non-trivial case is when a Liouvillian form $\theta:=\theta_\II=\theta_\JJ$ 
has linear components and the structure $\theta=\pi_1^*\theta_1 - \pi_2^*\theta_2$. 
In canonical coordinates $\{\bz_0,\bz_h\}$
it is necessary that 
$\theta_1 = d\bz_0\left( \fiz J_{2n} + \bS \right)\bz_0$
and $\theta_2 = d\bz_h\left( \fiz J_{2n} - \bS \right)\bz_h$, where $\bS\in\mathbb M_{2n\times 2n}(\mathbb R)$ 
is a symmetric Hamiltonian matrix. The symplectic map is given by 
$\bz_h= (I_{2n}-2\bb)^{-1}(I_{2n}+2\bb)\bz_0$, where $\bb=J_{2n}\bS$ is again  a symmetric Hamiltonian matrix. This symplectic map is just 
the solution to the equation $\pi_*(\II(v))=0$, where $v=Z(\bz_0,\bz_h)$ is the 
element of the Liouville vector field $Z$ in the point $(\bz_0,\bz_h)\in \mP$, 
and $Z$ is the dual of the Liouvillian form $\theta=i_Z\omega_\JJ$  \cite{Jim15a,Jim19a}.

For constructing symplectic integrators we use the Liouville vector field $Z$
for computing an intermediate point $\bar\bz$, where we must evaluate an Euler numerical 
scheme.
The intermediate point is given by $\bar\bz=\pi_*(v)$. For instance, 
given a \emph{Hamiltonian vector 
field} $X_H$ on $(M,\omega)$ 
with equations $\dot \bz = X_H(\bz)$,
a first order approximation of the flow of $\dot\bz$ is given by 
the implicit Euler scheme 
\begin{eqnarray}
    \bz_h = \bz_0 + h X_H(\bar \bz), \qquad \bar\bz,\bz_h,\bz_0\in M,\quad 0< h\ll 1.
    \label{eqn:ham}
\end{eqnarray}
This map is symplectic when the point $\bar \bz$ is given by \cite{KQ10,Jim15a} 
\begin{eqnarray}
  \bar\bz= \fiz(\bz_0 + \bz_h) + \bb(\bz_h - \bz_0).
   \label{eqn:cond}
\end{eqnarray}
It corresponds to the expression $\bar\bz=\pi_*(v)=\pi_*(Z(\bz_0,\bz_h))$. Moreover, 
the implicit Euler scheme is both symmetric and symplectic if 
$\bar\bz= \fiz(\bz_0 + \bz_h) + h\bb(\bz_h - \bz_0)$ \cite{Jim15a}.

All the one-step symplectic integrators are realizations of symplectic maps which arrive 
in this way. 
The three well-known one-step methods are the symplectic Euler methods $A$ and $B$ and the 
mid point rule. 
 The case $\bb=0_{2n}$ corresponds to 
the implicit mid-point $\bar\bz$ whose explicit symplectic map is the identity map. 
This is a degenerated case which corresponds to the flow of constant vector fields. 
On the other hand, the Euler methods have Hamiltonian matrices
\begin{eqnarray*}
    \bb_A = \fiz \left( 
     \begin{array}{c c}
       -I_{n}   & 0 \\
       0 &  I_n
    \end{array}
    \right),\qquad 
    \bb_B = \fiz \left( 
     \begin{array}{c c}
       I_n   & 0 \\
       0 &  -I_n
    \end{array}
    \right),\qquad \bb_A,\bb_B\in  \mathbb M_{2n\times 2n}(\mathbb R).
\end{eqnarray*} 

\begin{remark} $\bb_A$ and $\bb_B$ are exceptional matrices of rank $n$, \emph{i.e.}  
$\det(I_{2n}\pm 2\bb_a) = 0$ and they have not a well defined map of the 
form: $\bz_h= (I_{2n}-2\bb)^{-1}(I_{2n}+2\bb)\bz_0$. 
Their symplectic maps induce the only explicit 
one-step symplectic integrators. 
\end{remark}
The matrix $\bb$ is related with the curvature of the flow lines.
However, this subject is 
out of the scope of this paper.

\section{The Poincar\'e's Generating Function}
In the 3rd volume of \emph{Les m\'ethodes nouvelles de la m\'ecanique c\'eleste} \cite{Poi99}, 
Poincar\'e introduced the 1-form 
\begin{eqnarray}
    dS= \frac12\sum{ \left\{ (\bQ-\bq)d(\bP+\bp) - (\bP-\bp)d(\bQ+\bq)  \right\} }
    \label{eqn:poi}
\end{eqnarray}
looking for periodic orbits bifurcating from a prescribed periodic 
orbit of period $T>0$. In expression (\ref{eqn:poi}) variables $(\bq,\bp)$
are positions and conjugate momenta in the phase space at time $t$
and $(\bQ,\bP)$ are positions and conjugate momenta at time $t+T$.
We denote them by $\bz_0=(\bq,\bp)$ and $\bz_h=(\bQ,\bP)$. 
This form was rediscovered by Feng Kang and his collaborators when they were 
studying the construction of  symplectic integrators using generating functions. 
Kang's group  interprets the form (\ref{eqn:poi}) as the linear mapping 
\begin{eqnarray}
    \bp d\bq + \bP d\bQ \mapsto \fiz \left[ (\bQ-\bq)d(\bP+\bp) - (\bP-\bp)d(\bQ+\bq) \right]
    \label{eqn:dif}
\end{eqnarray}
given by the matrix
\begin{eqnarray}
    \alpha = \left( 
    \begin{array}{c c }
        -J_{2n} & J_{2n}\\
        \frac12 I_{2n} & \frac12 I_{2n}
    \end{array}
    \right),\qquad J_{2n},I_{2n}\in\mathbb M_{2n\times 2n}(\mathbb R).
    \label{eqn:alpha}
\end{eqnarray}
Given a generating function $u:\Lambda\to\mathbb R$ on a 
Lagrangian submanifold $\Lambda\subset \mP$ with local coordinates $w=w(\bz_0,\bz_h)$,
they systematically associate the numerical method 
\begin{eqnarray}
   \bz_h - \bz_0 = -J_0\frac{\partial u}{\partial w}\left( \frac{\bz_h+\bz_0}{2} \right).
   \label{eqn:feng:def}
\end{eqnarray}
to the form (\ref{eqn:poi}) \cite{Kang85a,Kang85b,KG88,KHMD89},
and consequently the symplectic mid-point rule with 
the Poincar\'e's generating  function.
Note that the map 
(\ref{eqn:feng:def}) is a time-1 map.

Poincar\'e's form (\ref{eqn:poi}) 
has linear components and it accepts a matrix representation $\theta = d\bx\left(\fiz \JJ + \bR\right)\bx^T=d\bx \bA\bx^T$, where
$\bx= (\bq,\bp,\bQ,\bP)=(\bz_0,\bz_h) \in\mP$, 
\begin{eqnarray}
    \bR = \fiz\left( 
    \begin{array}{c c }
        0_{2n} & J^T_{2n}\\
        J_{2n} & 0_{2n}
    \end{array}
    \right)
    \quad{\rm and}\quad
    \bA = \fiz\left( 
    \begin{array}{c c }
        J_{2n} & J^T_{2n}\\
        J_{2n} & J^T_{2n}
    \end{array}
    \right),\quad 
    J_{2n},I_{2n}\in\mathbb M_{2n\times 2n}(\mathbb R).
    \label{eqn:A}
\end{eqnarray}

In order to obtain the implicit Euler scheme we follow the same procedure in 
\cite{Jim19a}. Let $Z$ be the Liouville vector field for $\omega_\JJ$ 
dual to $\theta$ and we compute $v=Z(\bz_0,\bz_h)$. 
From (\ref{eqn:poi}) and (\ref{eqn:A}) we have, in local coordinates, 
$v=Z(\bz_0,\bz_h)= \JJ^T \bA \bx^T$. A direct computation shows that 
the point $\bar\bz$ is given by
\begin{eqnarray}
    \bar\bz =\pi_*(v) = \pi_*( \JJ\circ \bA \bx^T) =0_{2n}.
   \label{eqn:zbar}
 \end{eqnarray}
This is the point where we evaluate the Hamiltonian system and we consider that $\bar\bz = 0_{2n}$ 
is a fixed point $X_H(0_{2n}) = 0_{2n}$ as Poincar\'e did, obtaining 
\begin{eqnarray}
   \bz_h = \bz_0 + hX_H(0_{2n}) = \bz_0, 
\end{eqnarray}
which corresponds to the identity map for every $h\in\mathbb R$. 
Alternatively, the expression(\ref{eqn:poi}) defines a Lagrangian subspace 
by the equation $dS=0$ with solutions
$\bp-\bP=0$ and $\bQ - \bq=0$, which produce the identity map $\bP=\bp$ and $\bQ=\bq$. 
\begin{figure}
 \centering
 \includegraphics[scale=0.35]{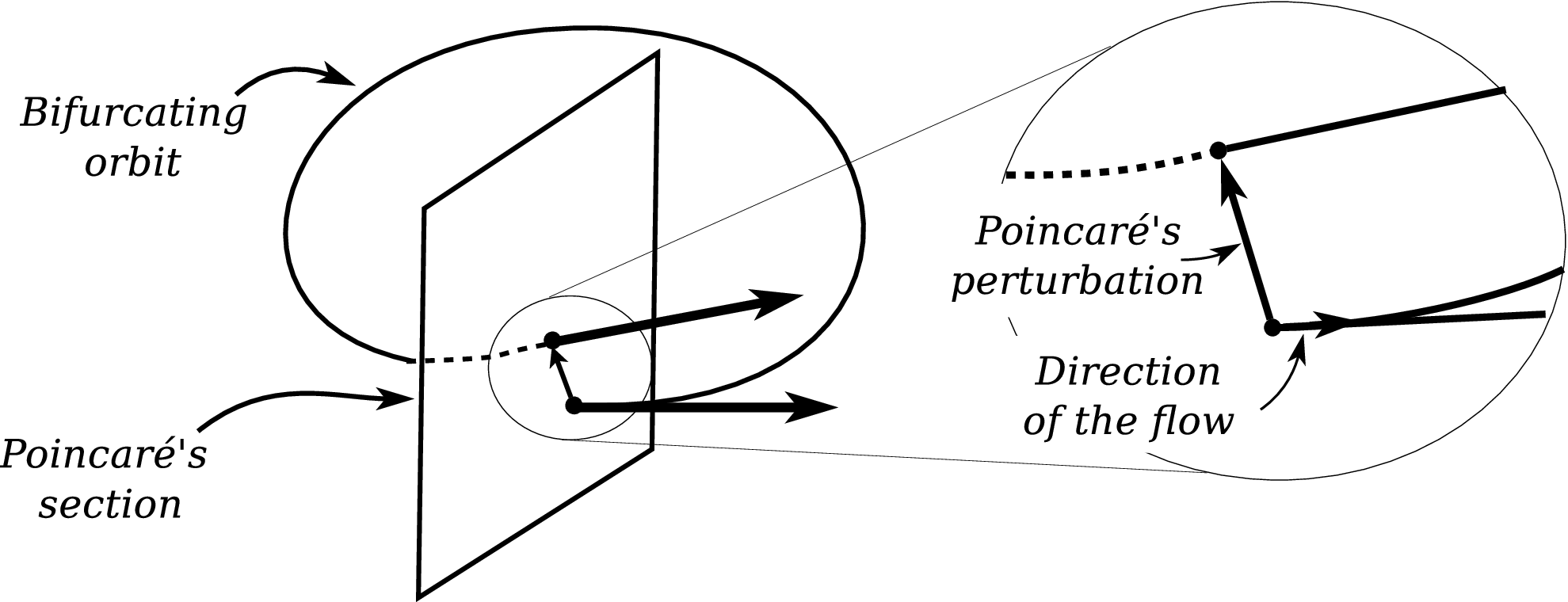}
 \caption{The original framework of Poincar\'e. The direction of the perturbation is perpendicular 
 to the flow's direction.}
 \label{fig:poin}
\end{figure}

This is nothing else that Poincar\'e's original framework, since he constructed 
the generating function to be defined on periodic orbits of prescribed period $T>0$ 
such that if $\bz_0=(\bq,\bp)$, $\bz_h=(\bQ,\bP)$ then $\bz_h=\bz_0$
was a fixed point on a Poincar\'e's section but such that the orbit was a non-trivial one 
(see Figure \ref{fig:poin}).
Moreover, this hypothesis considers that for $T$ variable, 
   $\lim_{T\to 0} dS = 0$
and the function $S$ goes to a constant $S\to S_0$ which, for simplicity, he considered $S_0=0$.
We have proven that our interpretation of the Poincar\'e's generating function using the framework of Liouvillain forms
matches with the original framework from Poincar\'e.  

Let $\theta:=\pi^*(dS)$ be the Liouvillian form obtained by the pull-back of (\ref{eqn:poi}).
Let $Z=\theta^\sharp$ be the Liouville vector field dual to $\theta$ under $\omega_\JJ$, \emph{i.e.}
$\theta=i_Z\omega_\JJ$, and let $v=Z(\bz_0,\bz_h)\in T_{(\bz_0,\bz_h)}\mP$ the corresponding vector of $Z$ at the point 
$(\bz_0,\bz_h)\in\mP$.
\begin{proposition}
   The implicit map $\bar\bz = \pi_*(v)$ associated to the Poincar\'e's form (\ref{eqn:poi}) under the method
   of Liouvillian forms corresponds to the null map (\ref{eqn:zbar}). If $\bar\bz$ is a fixed point of the 
   Hamiltonian vector field $X_H(\bar\bz) = \bar\bz$, 
   the generalized Euler scheme corresponds to the identity map.
\end{proposition}
%

%

It is well-known that symplectic maps close to the identity can be 
used to construct symplectic integrators \cite{HLW10}, but this is not the 
case for maps obtained by Poincar\'e's generating function. This is because 
the variational problem for which it was constructed assumes non-trivial 
periodic orbits with period $T>0$ \cite{Poi99}. If the fixed point is $\bar\bz\neq 0$
then it concerns a structurally different Liouvillian form as showed in the previous section. 
In fact, the Liouvillian form determines the evaluation point $\bar\bz$. We will show that 
the symplectic map produced by Poincar\'e's generating function is far from the set 
of symplectic maps producing regular symplectic integrators. For this, 
we construct a path of Liouvillian forms connecting
the symplectic Euler methods $A$ and $B$ with the linear form (\ref{eqn:poi}) using
a loop of symplectic rotations.

\begin{lemma}
    The 1-parameter family of Liouvillian forms on $(\mP,\omega_\JJ)$, given by
  \begin{eqnarray*}
      \theta_{\phi} &=& (\cos{\phi} \bQ - \sin{\phi}\bq )d(\cos{\phi} \bP + \sin\phi \bp) - (\sin{\phi} \bP - \cos{\phi}\bp )d(\sin{\phi} \bQ + \cos\phi \bq).
  \end{eqnarray*}
    connects Poincar\'e's 1-form,
    to those associated with the symplectic Euler schemes A and B.
    \label{prop:euler}
\end{lemma}
{\it Proof.} It is a family of Liouvillian forms on $(\mP,\omega_\JJ)$ since $d\theta_\phi=\omega_\JJ$ is independent of the 
parameter $\phi\in [0,2\pi]$. To prove that this family contains both Euler schemes and the 
Poincar\'e's 1-form, it is enough to compute $\theta_\phi$ 
for the values $\phi\in\{ 0, \pi/4, \pi/2\}$
obtaining 
\begin{eqnarray}
    \theta_{0} = \bp d\bq + \bQ d\bP,\qquad \theta_{\pi/2} = -\bq d\bp -\bP d\bQ,
    \label{eqn:lims1}
\end{eqnarray}
\begin{eqnarray}
    \theta_{\pi/4}= \fiz\left\{ (\bQ-\bq)d(\bP+\bp)-(\bP-\bp)d(\bQ+\bq) \right\}
    \label{eqn:lims2}
\end{eqnarray}
which corresponds to the forms associated with the symplectic Euler schemes $A$ and $B$ \cite{Jim15a} and the Poincar\'e's 1-form, 
respectively.
$\hfill\square$

\begin{figure}
 \centering
 \includegraphics[scale=0.22]{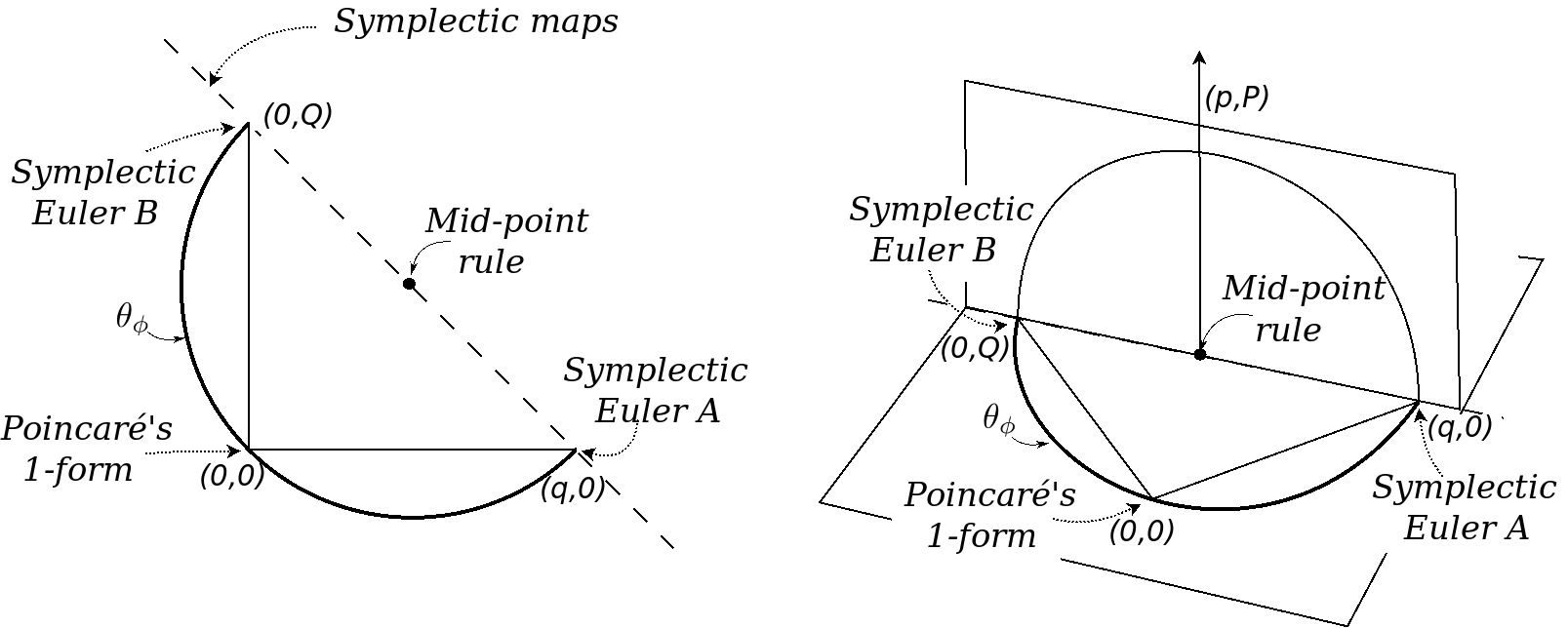}
 \caption{The family $\theta_\phi$ and its projection on the subspace 
     $(\mathcal Q_1\times\mathcal Q_2)$. 
    The set of Liouvillian forms whose projection gives symplectic integrators,
     reproduces the original positions and conjugated momenta on the diagonal 
 $(\bq=\bQ, \bp=\bP)$, which is not the case for the Poincar\'e's 1-form.}
 \label{fig:circle}
\end{figure}

Expand the family $\theta_\phi$ from Lemma \ref{prop:euler} in order to recover the matrix representation
of the family $\theta_\phi=d\bx \bA_\phi \bx^T$ where 
\begin{eqnarray*}
  \bA_\phi =\left( 
    \begin{array}{c c c c}
      0 & \cos^2\phi I_n & 0_n& -\cos\phi\sin\phi I_n\\
      -\sin^2\phi I_n & 0_n & \cos\phi\sin\phi I_n & 0_n\\
      0_n & \cos\phi\sin\phi I_n& 0_n & -\sin^2\phi I_n \\
      -\cos\phi\sin\phi I_n & 0_n & \cos^2\phi I_n& 0_n
    \end{array}
  \right). 
\end{eqnarray*}
Using trigonometric identities 
we rewrite $\bA_\phi =\fiz\left(  \JJ + \cos2\phi\ \bS_c +  \sin2\phi\ \bS_s\right)$,
where 
\begin{eqnarray*}
  \bS_c =
 \left( 
    \begin{array}{c c c c}
      0_n & I_n & 0_n& 0_n\\
      I_n & 0_n & 0_n & 0_n\\
      0_n & 0_n& 0_n & I_n \\
      0_n & 0_n & I_n& 0_n
    \end{array}
  \right)
  \qquad{\rm and} \qquad
  \bS_s =\left( 
    \begin{array}{c c}
      0_{2n} & J^T_{2n} \\
      J_{2n} & 0_{2n} 
    \end{array}
  \right).
\end{eqnarray*}
Elements of the family $\theta_\phi$ have the shape $\theta_\ominus = \pi_1^*\theta_1 - \pi_2^*\theta_2$
when the matrix $\bA_\phi$ is block diagonal, and it happens if and only if $\sin 2\phi=0$.
This condition is satisfied in 
$[0,\pi/2]$, for $\phi=0$ or $\phi=\pi/2$.
Consequently the Hamiltonian matrices $\bb_A= \JJ^T \bS_c$ and $\bb_B= \JJ \bS_c$
are block diagonal matrices 
corresponding to the matrices of the symplectic Euler 
schemes $A$ and $B$ (see Figure \ref{fig:circle}),
we have proven the following:
\begin{theorem}
   The family of Liouvillian forms $\theta_\phi$ from Lemma \ref{prop:euler} renders the generalized implicit Euler method,
   symplectic if and only if 
   $\phi=0$ or $\phi=\pi/2$, equivalently  if and only if it is one of the symplectic Euler schemes: A or B.
   \label{teo:1}
\end{theorem}

Then we cannot construct a well defined symplectic integrator by this formalism using
Poincar\'e's generating function. 
This negative result has a simple explanation in the framework of Liouvillian forms:
the Liouville vector field induced by Poincar\'e's generating function (\ref{eqn:poi})
is Liouville for $\omega_\JJ$ but it is not Liouville for $\omega_\II$. As a consequence, 
the submanifold $\Lambda\stackrel{\jmath}{\hookrightarrow} \mP$ associated to $dS$ is Lagrangian with respect to 
$\omega_\JJ$ but not for $\omega_\II$ nor $\omega_\KK$.

\section{The Liouvillian form for the mid point rule}

Looking for the most generic Liouvillian form on $(\mP,\omega_\JJ)$ which induces the 
mid point rule as discrete map, we consider the expression for $\bar\bz$ corresponding to
$$\bar \bz = \pi_*( v ) = \pi_* \left\{\left(\fiz I_{4n} + \JJ^T\bS\right)\bx^T\right\} \equiv \fiz(\bz_0+\bz_h),$$
where $\bx = (\bz_0,\bz_h)\in \mP$ and $\bS\in\mathbb M_{4n\times 4n}(\mathbb R)$ is a symmetric matrix.
This expression is satisfied when $\pi_*\left(\JJ^T\bS\bx^T\right) =0_{2n}$. 
In behalf of simplicity, we consider $\mP=\mathbb R^{4n}$ for avoiding curvature issues, and 
we write $\bS$ in $2n\times 2n$ blocks. We have 
\begin{eqnarray*}
  \bS =
 \left( 
    \begin{array}{c c}
      \bS_1 & \bG_1 \\
      \bG_1^T & \bS_2
    \end{array}
  \right) \quad\Rightarrow \quad
  \JJ^T\bS\bx^T=
  \left(\JJ^T \bS\right)
  \left(\begin{array}{c}
         \bz_0\\
         \bz_h
        \end{array}
        \right)=
 \left( 
    \begin{array}{c}
      J^T_{2n}\bS_1\bz_0 + J^T_{2n}\bG_1^T\bz_h \\
      J_{2n}\bG_1\bz_0 + J_{2n}\bS_2\bz_h  
    \end{array}
  \right)
\end{eqnarray*}
where $\bS_1$ and $\bS_2$ are symmetric matrices and $\bG_1$ is a generic matrix in $\mathbb M_{2n\times 2n}(\mathbb R)$. 
Condition $\pi_*\left(\JJ^T\bS\bx^T\right) =0_{2n}$ holds
if and only if $\bG_1=\bS_1=\bS_2$, and the matrix $\bS$ has the form
\begin{eqnarray*}
  \bS =
 \left( 
    \begin{array}{c c}
      \bS_1 & \bS_1 \\
      \bS_1 & \bS_1
    \end{array}
  \right),\qquad \bS_1=\bS_1^T\in \mathbb M_{2n\times 2n}(\mathbb R).
\end{eqnarray*}
This corresponds to a family of $n(2n+1)$ free parameters producing the mid-point rule,
which accepts a description in the form $\theta= \pi_1^*\theta_1 - \pi_2^*\theta_2$
if and only if $\bS_1\equiv 0_{2n}$, since
$\bS\in \mathbb M_{4n\times 4n}(\mathbb R)$ must be a block diagonal matrix.

%

\begin{theorem}
The Liouvillian form associated to the mid-point rule has a unique element of the type 
$\theta = \pi^*_1\theta_1-\pi^*_2\theta_2$ corresponding to the basic Liouvillian form 
$\theta_0$. In local coordinates $\bx\in(\mP,\omega_\JJ)$ we write $\theta_0= \fiz d\bx \JJ \bx$
or in extended form
  \begin{eqnarray}
      \theta_{0} &=& \fiz(\bp d\bq - \bq d\bp - \bP d\bQ + \bQ d\bP),\qquad \bx=(\bq,\bQ,\bp,\bP)\in(\bP,\omega_\JJ).
      \label{eqn:mid:point2}
  \end{eqnarray}

\end{theorem}

Moreover, the Liouville vector field $Z$ associated to the basic Liouvillian form $\theta_0$ is just 
the ``expanding'' or Euler vector field. If we consider local coordinates $\{x_i\}_{i=0}^{4n}$
then $Z=\fiz\sum_ix_i\frac{\partial}{\partial x_i}$ is Liouville for 
the three symplectic forms $\omega_\II$, $\omega_\JJ$ and $\omega_\KK$. 
This also implies that the mid-point rule is a degenerated case of Liouvillian forms 
for constructing symplectic maps, corresponding to the flow of constant 
(Hamiltonian) vector fields. This comes from the expression $\bz_h = (I_{2n} - 2\bb)^{-1}(I_{2n} + 2\bb)\bz_0$
since the mid-point rule corresponds to $\bb=0_{2n}$. 

The degeneracy is related with the dimension of the immersed submanifold $\Lambda$. 
Fixing the almost quaternionic structure  $\{I_{4n},\II,\JJ,\KK\}$ on $\mP$, 
an immersion $\jmath:\Lambda\hookrightarrow\mP$ which
is Lagrangian with respect to $\omega_\II$ and $\omega_\JJ$, must be symplectic 
with respect to $\omega_\KK$. In the case of the mid-point rule, the Liouville vector fields 
for the different symplectic forms coincide, and consequently the immersion corresponds to 
an isotropic submanifold. 

The argument for naming  $\theta_0$ the \emph{basic Liouvillian form} is based on the Hodge decomposition of 
differential forms on a differential manifold \cite{Mor01}. In this decomposition, 
every Liouvillian form on a symplectic manifold $(M,\omega)$ is given by $\theta=\delta\eta+dF+ \alpha$, 
where $\alpha$ is a harmonic form, $dF$ is the differential of a function $F:M\to\mathbb R$, and $\delta\eta$
is the codifferential of a 2-form $\eta\in\Omega^2(M)$. Then $\theta_0=\delta\eta$ is the 
only contribution to the symplectic form $\omega=d\theta_0 = d\delta\eta$ since $d(dF+h)=0$.
It is a different point of view than the geometrical interpretation of the Liouville 
form as a tautological form on a cotangent bundle.

\section {Conclusions}
In this paper we used Liouvillian forms \cite{Jim15a}
for studying the relation of the mid-point rule with Poincar\'e's 1-form, which is the differential of the Poincare's generating 
function  (\ref{eqn:poi}), introduced in 
\cite{Poi99}. We showed that the classical association 
between these two objects is not the right one. This comes from the fact that 
Poincar\'e's 1-form and mid-point rule are techniques applied to two different 
types of variational problems:
\begin{itemize}
   \item Poincar\'e's 1-form was designed for dealing with  periodic orbits with 
   prescribed period $T>0$, it means, non-trivial loops or cycles (no boundary);
   \item the mid-point rule is the simplest approximation for problems 
   with fixed values at the boundary (initial and final fixed points).
\end{itemize}

We showed that the structure of Poincar\'e's 1-form 
differs drastically from the structure of Liouvillian forms 
generating the mid-point rule, and in general to those generating
symplectic integrators. 
In order to better understand this discrepancy, we constructed two 
families of 1-forms. The first one is a one-parameter family (a path)
joinning the symplectic Euler maps $A$ and $B$ with Poincar\'e's 1-form. The only elements in this 
family which generate symplectic integrators are the boundary points 
of the path corresponding to the Euler maps $A$ and $B$. 
The second family shows that the only Liouvillian form of type $\theta_\ominus=\pi^*_1\theta_1-\pi^*_2\theta_2$ producing the mid-point rule 
on the product manifold, is the basic Liouvillian form $\theta_0$ which has null symmetric part. 

\section*{Acknowledgements}
This research was developed with support from the Fondation du Coll\`ege de 
France and Total under the research convention PU14150472, as well as the ERC Advanced Grant 
WAVETOMO, RCN 99285, Subpanel PE10 in the F7 framework.

\end{document}